\documentclass[floatfix,aps,prl,twocolumn,superscriptaddress]{revtex4-2}
\usepackage[dvipdfmx]{graphicx}
\usepackage[]{epsfig}
\usepackage{textcomp}
\usepackage{times,amsmath,amssymb, amsfonts}
\usepackage[T1]{fontenc}
\usepackage{braket}
\usepackage{upgreek}
\usepackage{color}
\usepackage[colorlinks = true,
            linkcolor = blue,
            urlcolor  = blue,
            citecolor = blue,
            anchorcolor = blue]{hyperref}

\begin{document}

\title{Charge sensing of few-electron ZnO double quantum dots probed by radio-frequency reflectometry}

\author{Kosuke Noro}
\affiliation{Research Institute of Electrical Communication, Tohoku University, 2-1-1 Katahira, Aoba-ku, Sendai 980-8577, Japan}
\affiliation{Department of Electronic Engineering, Graduate School of Engineering, Tohoku University, 6-6 Aramaki Aza Aoba, Aoba-ku, Sendai 980-0845, Japan}

\author{Motoya Shinozaki}
\affiliation{WPI Advanced Institute for Materials Research, Tohoku University, 2-1-1 Katahira, Aoba-ku, Sendai 980-8577, Japan}

\author{Yusuke Kozuka}
\affiliation{Research Center for Materials Nanoarchitechtonics (MANA), National Institute for Material Science (NIMS),
1-2-1 Sengen, Tsukuba 305-0047, Japan}

\author{Kazuma Matsumura}
\affiliation{Research Institute of Electrical Communication, Tohoku University, 2-1-1 Katahira, Aoba-ku, Sendai 980-8577, Japan}
\affiliation{Department of Electronic Engineering, Graduate School of Engineering, Tohoku University, 6-6 Aramaki Aza Aoba, Aoba-ku, Sendai 980-0845, Japan}

\author{Yoshihiro Fujiwara}
\affiliation{Research Institute of Electrical Communication, Tohoku University, 2-1-1 Katahira, Aoba-ku, Sendai 980-8577, Japan}
\affiliation{Department of Electronic Engineering, Graduate School of Engineering, Tohoku University, 6-6 Aramaki Aza Aoba, Aoba-ku, Sendai 980-0845, Japan}

\author{Takeshi Kumasaka}
\affiliation{Research Institute of Electrical Communication, Tohoku University, 2-1-1 Katahira, Aoba-ku, Sendai 980-8577, Japan}

\author{Atsushi Tsukazaki}
\affiliation{Department of Applied Physics and Quantum-Phase Electronics Center (QPEC), University of Tokyo, 7-3-1 Hongo, Bunkyo-ku, Tokyo 113-8656, Japan}

\author{Masashi Kawasaki}
\affiliation{Department of Applied Physics and Quantum-Phase Electronics Center (QPEC), University of Tokyo, 7-3-1 Hongo, Bunkyo-ku, Tokyo 113-8656, Japan}
\affiliation{Center for Emergent Matter Science, RIKEN, 2-1 Hirosawa, Wako, Saitama 351-0198, Japan}

\author{Tomohiro Otsuka}
\email[]{tomohiro.otsuka@tohoku.ac.jp}
\affiliation{Research Institute of Electrical Communication, Tohoku University, 2-1-1 Katahira, Aoba-ku, Sendai 980-8577, Japan}
\affiliation{Department of Electronic Engineering, Graduate School of Engineering, Tohoku University, 6-6 Aramaki Aza Aoba, Aoba-ku, Sendai 980-0845, Japan}
\affiliation{WPI Advanced Institute for Materials Research, Tohoku University, 2-1-1 Katahira, Aoba-ku, Sendai 980-8577, Japan}
\affiliation{Center for Science and Innovation in Spintronics, Tohoku University, 2-1-1 Katahira, Aoba-ku, Sendai 980-8577, Japan}
\affiliation{Center for Emergent Matter Science, RIKEN, 2-1 Hirosawa, Wako, Saitama 351-0198, Japan}

\date{\today}

\begin{abstract}
Zinc oxide (ZnO) has garnered much attention as a promising material for quantum devices due to its unique characteristics.
To utilize the potential of ZnO for quantum devices, the development of fundamental technological elements such as high-speed readout and charge sensing capabilities has become essential. 
In this study, we address these challenges by demonstrating radio-frequency (rf) reflectometry and charge sensing in ZnO quantum dots, thus advancing the potential for qubit applications.
A device is fabricated on a high-quality ZnO heterostructure, featuring gate-defined target and sensor quantum dots. 
The sensor dot, integrated into an rf resonator circuit, enables the detection of single-electron charges in the target dots. 
Using this setup, the formation of few-electron double quantum dots is observed by obtaining their charge stability diagram.
Also, a charge stability diagram with a gate pulse sequence is measured.
We discuss the strong electron correlation in ZnO, which leads to nearly degenerate spin-singlet and -triplet two-electron states in the (0, 2) charge state, and the perspectives on spin-state readout.
\end{abstract}

\maketitle

Semiconductor quantum dots, called artificial atoms, are representative systems that allow the control of quantum states~\cite{tarucha1996shell, kouwenhoven1997excitation, kouwenhoven2001few}.
Few-electron quantum dots are used as quantum bits (qubits), which are the fundamental building blocks of quantum computing~\cite{maurand2016cmos, zwerver2022qubits}.
Many efforts have been underway to develop qubits with one approach based on a deep understanding of material engineering~\cite{kanai2022generalized}.
Consequently, materials are now being conducted not only on conventional semiconductors such as gallium arsenide (GaAs)~\cite{koppens2006driven, yoneda2014fast} but also on isotopically controlled silicon (Si)~\cite{fogarty2018integrated, yoneda2018quantum, takeda2022quantum, noiri2022fast}, two-dimensional materials like graphene~\cite{banszerus2021dispersive, johmen2023radio, garreis2024long, tataka2024surface, ruckriegel2024electric}, and so on.
These materials have advanced qubit characteristics, particularly with respect to the coherence time of quantum states.
Zinc oxide (ZnO) is also considered a promising candidate owing to its low nuclear spin density, high electron mobility~\cite{ falson2015electron,falson2016mgzno}, and direct bandgap transition~\cite{tsukazaki2005blue}, which is advantageous to long coherence times~\cite{kanai2022generalized,Linpeng2018coherence,Niaouris2022ensemble} and coupling with photons~\cite{Yan2023}.
Quantum transport phenomena such as the quantum Hall effect~\cite{tsukazaki2007quantum,tsukazaki2010observation,falson2015even}, quantum point contacts~\cite{hou2019quantized}, and notably quantum dots~\cite{Noro2024} have been reported, leading to increasing research on quantum devices utilizing high-quality ZnO heterostructures.
However, qubit operation in ZnO remains challenging, as it requires the formation of few-electron states in quantum dots~\cite{tarucha1996shell, kouwenhoven1997excitation, kouwenhoven2001few} and the development of high-speed qubit readout techniques.

One of the challenges to realize few-electron states is to probe the number of electrons in the quantum dot. This is because the source-drain current across the dot cannot always be used to probe the quantum dot states as a gate voltage to tune the energy of the dot often suppresses the tunnel current between the dot and the lead electrodes~\cite{sprinzak2002charge}. 
Therefore, an alternative measurement method is expected to accurately count the number of electrons in gate-defined quantum dots.
To address these issues, a quantum point contact or another quantum dot has been used as a charge sensor to accurately probe the number of electrons in the quantum dot in focus~\cite{field1993measurements, elzerman2003few, kurzmann2019charge}. Additionally, charge sensing using radio-frequency (rf) reflectometry is a powerful tool, capable of detecting a single electron charge with a broad bandwidth~\cite{qin2006radio, reilly2007fast, barthel2009rapid}.
In this method, the sensor's conductance is reflected in the reflection coefficient of the rf signal applied to the resonator, including the sensor dot, and its reflection coefficient depends on the sensor's conductance.
This technique avoids parasitic capacitance in the circuit line and low-frequency noise, which limits the bandwidth of direct current measurements, enabling fast readout of quantum dynamics.

In this paper, we fabricate a device structure using a ZnO heterostructure to form target and sensor quantum dots defined by gate voltage.
The sensor dot is integrated into the resonator circuit to demonstrate rf reflectometry and provides a charge stability diagram of few-electron double quantum dots. We also discuss the effect of strong electron correlation on the spin blockade measurement. 

\begin{figure}
\begin{center}
  \includegraphics{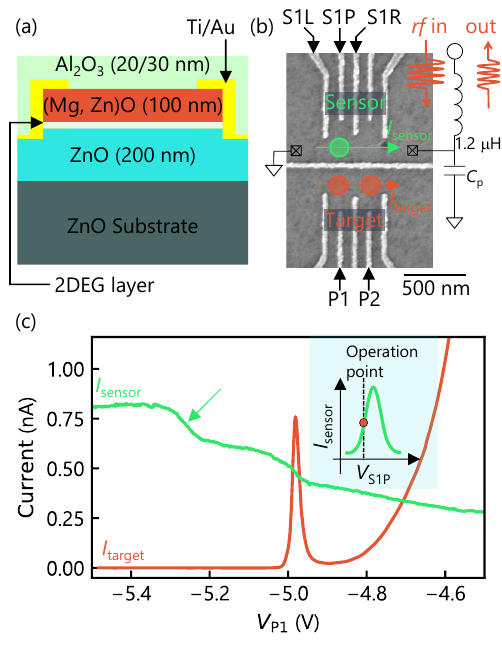}
  \caption{(a) Schematic of the device structure. 
(b) Scanning electron microscope image of the fabricated device and the resonance circuit.
(c) Gate voltage $V_\mathrm{P1}$ dependence of the sensor current $I_\mathrm{sensor}$ and target current $I_\mathrm{target}$. The inset shows an operation point of the sensor dot.}
  \label{device}
\end{center}
\end{figure}

The layer structure of the device is illustrated in Figure~\ref{device}(a).
The numbers in parentheses represent the layer thicknesses in nanometers.
The (Mg,Zn)O/ZnO heterostructure is grown on a Zn-polar ZnO (0001) substrate using molecular beam epitaxy.
Ti/Au ohmic and gate electrodes are fabricated through photolithography and electron beam lithography, respectively, followed by a lift-off process.
An AlO$_\mathrm{x}$ gate insulator is deposited using atomic layer deposition.
We prepare two types of stack structures: one with a 20-nm AlO$_\mathrm{x}$ layer and another with a 30-nm layer. 
The latter structure is used for spin-state readout measurements.
As shown in Fig.~\ref{device}a, the two-dimensional electron gas forms at the interface between the (Mg,Zn)O and ZnO layers.
The Mg content in the heterostructure is approximately 2.5 \%. 
The electron density and mobility are measured by the Hall effect as $n = 4.9\times10^{11}$~cm$^{-2}$ and $\mu = 170,000$~cm$^{2}$ V$^{-1}$ s$^{-1}$, respectively, at 1.8~K.

Figure~\ref{device}(b) shows a scanning electron microscope image of the device and the schematic resonator.
The upper dots correspond to the charge sensor, while the bottom to the target quantum dots. 
The sensor quantum dot is embedded in an rf resonator circuit constructed with a 1.2 $\upmu$H chip inductor and stray capacitance $C_\mathrm{p}$.
All experimental measurements are performed within a dilution refrigerator maintained at 60 mK.

To demonstrate the charge sensor operation, we measure two types of source-drain current: one through the sensor dot $I_\mathrm{sensor}$ and another through the target dot $I_\mathrm{target}$.
Figure~\ref{device}(c) shows $I_\mathrm{sensor}$ and $I_\mathrm{target}$ as a function of the gate voltage $V_\mathrm{P1}$, where
the sensor's operating gate voltage $V_\mathrm{S1P}$ is set to the slope of the Coulomb peak as shown in the inset of Fig.~\ref{device}c.
$I_\mathrm{target}$ exhibits a typical Coulomb peak behavior, indicating the formation of a quantum dot.
At the peak position of $I_\mathrm{target}$ around $V_\mathrm{P1}=-5.0$ V, the step-like variation of $I_\mathrm{sensor}$ reflects the increase or decrease of electrons in the target dot because the electron acts as an effective gate voltage on the sensor dot through the electrostatic coupling between the target and sensor dots.
Note that $V_\mathrm{S1P}$ is compensated to maintain the operation point, as $V_\mathrm{P1}$ affects the sensor dot conductance.
Another noticeable change in $I_{\mathrm{sensor}}$ is observed at $V_\mathrm{P1} \approx -5.25$ V (indicated by the arrow) while $I_\mathrm{target}$ remains nearly constant at zero.
This is because, in this region, the tunnel couplings between the target dot and the lead electrodes are extremely weak, resulting in nearly zero current, even when an electron is added to the target dot.
This behavior is also observed in the previous study in GaAs by using a quantum point contact sensor~\cite{reilly2007fast}.
These results demonstrate that the charge sensor enables accurate counting of the number of electrons in the target dot, even when the direct current through the target dot is too small to be detected.

\begin{figure}
\begin{center}
  \includegraphics{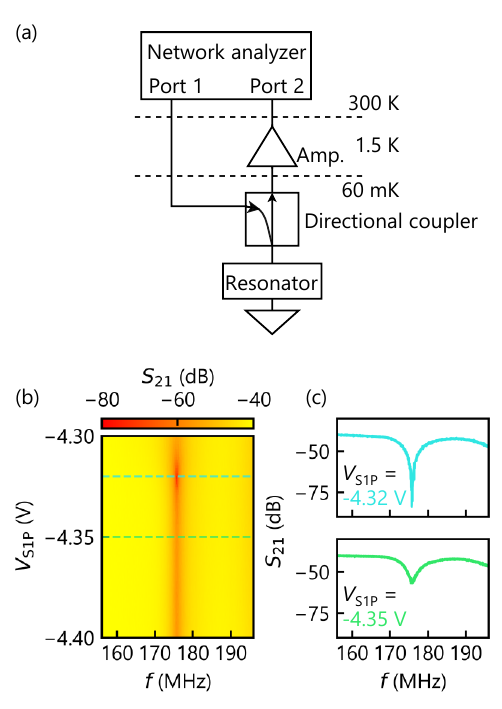}
  \caption{(a) Measurement setup to evaluate the reflection coefficients.
  (b) $V_\mathrm{S1P}$ dependence of the resonator's $S_{21}$ parameter. The color scale represents the magnitude of $S_{21}$ in dB.
  (c) $S_{21}$ characteristics showing typical traces with ($V_\mathrm{S1P}=-4.32$ V) and without ($V_\mathrm{S1P}=-4.35$ V) impedance matching.}
  \label{S21}
\end{center}
\end{figure}

Next, we investigate the rf-circuit properties to perform the rf reflectometry.
The reflected signal from the resonator is measured using a network analyzer connected through a directional coupler as shown in Fig.~\ref{S21}(a).
Figure~\ref{S21}(b) shows the transmission coefficients $S_{21}$ as functions of frequency and $V_\mathrm{S1P}$, while maintaining the gate voltages $V_\mathrm{S1R}$ and $V_\mathrm{S1L}$ at fixed values of $-3.84$ V and $-4.42$ V, respectively.
Note that $S_{21}$ is determined by measuring the input signal and the reflected signal from the resonator using a directional coupler, effectively representing the reflection coefficients.
The resonance frequency is approximately $174$ MHz, corresponding to $C_\mathrm{p}$ of $0.69$ pF.
The dip depth is modulated by $V_\mathrm{S1P}$, reflecting the changes in conductance of the sensor dot, while the resonance frequency remains nearly constant.
Figure~\ref{S21}(c) shows typical traces of $S_{21}$.
At $V_\mathrm{S1P} = -4.32$ V, a pronounced dip is observed where the impedance matching condition is satisfied.
In contrast, at $V_\mathrm{S1P} = -4.35$ V, a trace without impedance matching shows a small dip depth.
This large variation in dip depth enables high-sensitivity charge sensing under matching conditions.

\begin{figure*}
\begin{center}
  \includegraphics{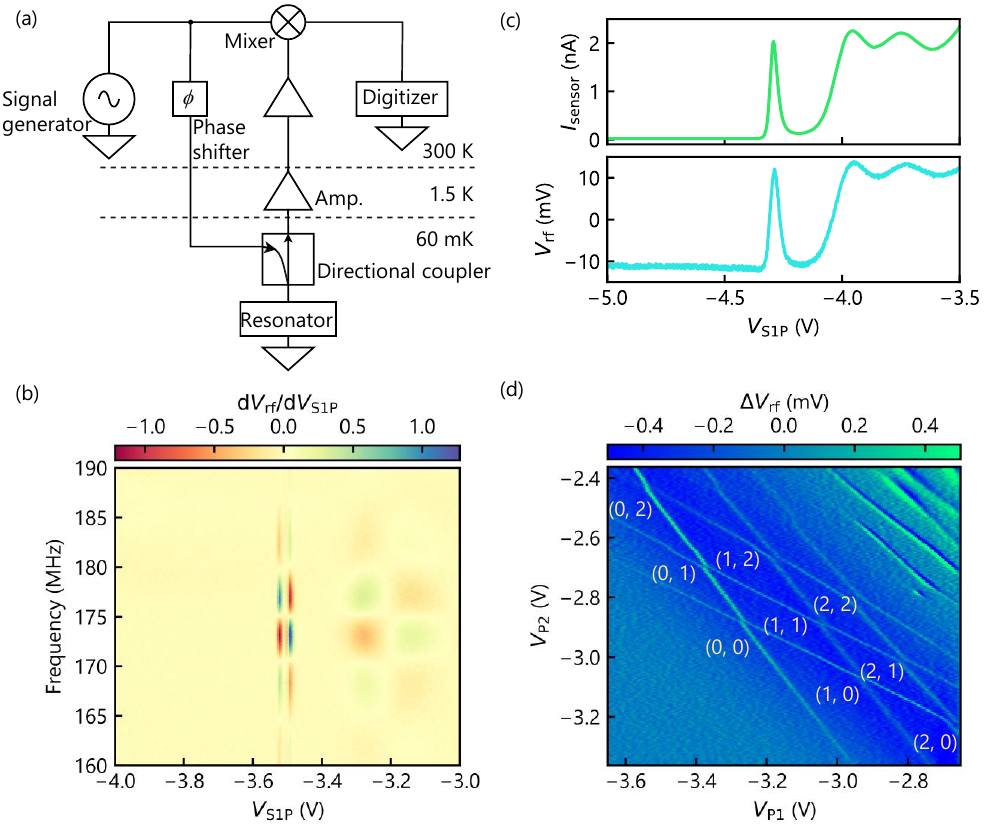}
  \caption{ (a) Schematic of the demodulation circuit.
(b) Frequency and $V_\mathrm{S1P}$ dependence of $\mathrm{d}V_\mathrm{rf}/\mathrm{d}V_\mathrm{S1P}$.
(c) $I_\mathrm{sensor}$ and $V_\mathrm{rf}$ as a function of $V_\mathrm{S1P}$.
(d) Charge stability diagram of a few electron double quantum dots probed by $V_\mathrm{rf}$.
$\Delta V_\mathrm{rf}$ is the numerical difference in $V_\mathrm{rf}$ along the $V_\mathrm{P1}$ axis direction.
}
  \label{RF_Sensor}
\end{center}
\end{figure*}

Figure~\ref{RF_Sensor}(a) illustrates a demodulation circuit for performing rf reflectometry. 
The resonator, which incorporates a charge sensor, receives an rf signal at a frequency of $174$ MHz via a directional coupler.
The amplitude of the rf signal is set to $-88$ dBm at the sample end.
After reflection, this rf signal undergoes amplification and is then mixed with a local oscillator signal for demodulation.
We optimize the phase shifter between the signal generator and the mixer.
The resulting down-converted signal voltage, denoted as $V_\mathrm{rf}$, is digitized.
The sampling rate is set to $100$ MHz.
To find the operation point, we measure the frequency and $V_\mathrm{S1P}$ dependence of a numerical derivative $\mathrm{d}V_\mathrm{rf}/\mathrm{d}V_\mathrm{S1P}$, as illustrated in Figure~\ref{RF_Sensor}(b).
A phase difference between the reflected signal and the local oscillator at the mixer modulates $V_\mathrm{rf}$.
A typical additional phase is caused by a phase shift induced by the circuit line, whose effects appear as periodic changes on $V_\mathrm{rf}$ along the frequency axis.
This color pattern, also observed in previous studies using GaAs and GaN quantum dots~\cite{shinozaki2021gate, fujiwara2023wide}, shows optimum operation points where the maximum sensitivity is obtained.

Figure~\ref{RF_Sensor}(c) shows $I_\mathrm{sensor}$ and $V_\mathrm{rf}$ with sweeping $V_\mathrm{S1P}$, where
we perform an integration of $2\time10^5$ data points to reduce the noise in $V_\mathrm{rf}$.
The changes in $V_\mathrm{rf}$ and $I_\mathrm{sensor}$ are synchronized, allowing us to detect the change in $I_\mathrm{sensor}$ by monitoring $V_\mathrm{rf}$.
The region where $V_\mathrm{rf}$ approaches zero satisfies the impedance matching condition, enabling high sensitivity for rf charge sensing.
After establishing the rf charge sensing technique, we apply this technique to probe the formation of double quantum dots in the target region.
The charge stability diagram obtained by monitoring $V_\mathrm{rf}$ while sweeping $V_\mathrm{P1}$ and $V_\mathrm{P2}$ is shown in Fig.~\ref{RF_Sensor}(d).
Two distinct slopes in the charge transition lines indicate the formation of double quantum dots.
The absence of charge transition lines in the more negative gate voltage region indicates the full depletion of the double quantum dot. 
By counting the number of electrons using the charge transition lines, we can assign the number of electrons in each region as shown in the figure. 
This result clearly shows we achieve few-electron states in this double quantum dot system.

\begin{figure}
\begin{center}
  \includegraphics{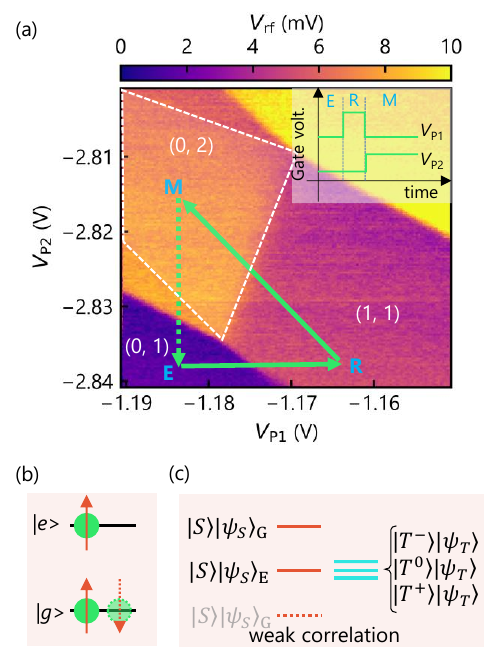}
  \caption{(a) Charge stability diagram with a gate pulse sequence under an in-plane magnetic field of $50$ mT. The inset indicates a three-step gate pulse sequence to observe the spin blockade.
(b) Energy levels of the ground and excited orbital states. With strong electron correlation, the formation of the ground singlet is forbidden and the higher orbital is occupied by the second electron.
(c) Energy levels of two-electron states considering both orbital and spin configurations with strong electron correlation. The dotted line indicates the ground state in the case of weak electron correlation.
  }
  \label{SB}
\end{center}
\end{figure}
With the few-electron double quantum dots realized, we can proceed to spin-state readout measurements.
For reading out the spin states utilizing the spin blockade measurement~\cite{ono2002current, koppens2005control}, we measure the stability diagram with a gate voltage pulse sequence under an in-plane magnetic field of $50$ mT~\cite{johnson2005triplet, barthel2009rapid, noiri2020radio}.
Here, the sampling rate is set to 125 MHz.
We perform this measurement near the (0,2) to (1,1) charge transition, where ($n_\mathrm{P1}$, $n_\mathrm{P2}$) indicates the number of electrons in the quantum dots defined by $V_\mathrm{P1}$ and $V_\mathrm{P2}$.
At the reset pulse step R, the two-electron state is initialized to either a singlet or one of the three triplet states.
If initialized to the singlet state, the state is transferred to the (0,2) charge state during the next measure pulse step M.
Conversely, if initialized to one of the triplet states, it remains in the (1,1) charge state during this pulse M.
After the measurement, one of the electrons is evacuated during the pulse step E to prepare the next reset phase.
As a result, the signal from the (1,1) state is expected to appear in the region near the boundary between the (0,2) and (1,1) states, approximately within the dashed region in Fig.~\ref{SB}(a)~\cite{johnson2005triplet, barthel2009rapid, noiri2020radio}.
However, the absence of signals in these regions indicates that spin blockade does not occur and/or the spin state might be relaxed by not fully optimized tunneling rates. 
Note that this measurement is performed in a different device with similar structures.

As to a possible mechanism, we will discuss the effect of electron correlation in ZnO.
For a two-electron state, considering both orbital and spin degrees of freedom, the spin states almost degenerate in the low magnetic field. 
These spin states are represented as $\ket{T^+}=\ket{\uparrow}_1\ket{\uparrow}_2$, $\ket{T^0}=\frac{1}{\sqrt{2}}\left(\ket{\uparrow}_1\ket{\downarrow}_2+\ket{\downarrow}_1\ket{\uparrow}_2\right)$, $\ket{T^-}=\ket{\downarrow}_1\ket{\downarrow}_2$, and $\ket{S}=\frac{1}{\sqrt{2}}\left(\ket{\uparrow}_1\ket{\downarrow}_2-\ket{\downarrow}_1\ket{\uparrow}_2\right)$. 
As a result, the energy of the electronic state is determined by the orbital energy.
Regarding the (0, 2) state, the orbital states for the ground (excited) spin-singlet $\ket{\psi_S}_\mathrm{G} (\ket{\psi_S}_\mathrm{E})$ and triplet $\ket{\psi_T}$ configurations can be expressed as
\begin{equation}
\begin{split}
&\ket{\psi_S}_\mathrm{G}=\ket{g}_1\ket{g}_2, \\
&\ket{\psi_S}_\mathrm{E}=\frac{1}{\sqrt{2}}\left(\ket{g}_1\ket{e}_2+\ket{e}_1\ket{g}_2\right),
\label{eq1}
\end{split}
\end{equation}
\begin{equation}
\ket{\psi_T}=\frac{1}{\sqrt{2}}\left(\ket{g}_1\ket{e}_2-\ket{e}_1\ket{g}_2\right).
\label{eq2}
\end{equation}
Here, $\ket{g}$ and $\ket{e}$ are the ground and excited orbital states, as shown in Fig.~\ref{SB}(b).
If the Coulomb interaction energy is larger than the difference between the ground and excited orbital levels, a second electron added to the quantum dot occupies the excited level.
Under this condition, $\ket{\psi_S}_\mathrm{E}$ becomes the ground state, whereas in devices with weak electron correlation, such as GaAs, $\ket{\psi_S}_\mathrm{G}$ typically assumes this role.
Consequently, even in the (0, 2) state, two-electron states with different spin configurations are nearly degenerate in energy, as shown in Fig.~\ref{SB}(c).
This is in contrast to the usual scenario where only the singlet state forms the ground state $\ket{S}\ket{\psi_S}_\mathrm{G}$, enabling readout operations by spin-blockade.
Indeed, the Kondo effect has been observed even in even-electron states in the ZnO quantum dot as a signature of strong electron correlation~\cite{maryenko2021interplay,Noro2024}.
Such behavior based on strong correlation supports the presented physical pictures, and our results suggest that improvements to the current situation or the establishment of new principles for readout methods are necessary.
One possible approach to the first solution is fabricating devices with smaller designs, increasing the orbital energy spacing than the electron interaction.

In this study, we demonstrate the charge sensing and rf reflectometry of ZnO quantum dots.
The sensor dot detects the charge state of the target dots, which has been difficult to determine when measuring the source-drain current of the target directly.
Owing to the charge sensor, we observe the few-electron regime of a double quantum dot system, which is essential for utilizing these systems as qubits.
Additionally, we discuss the spin state readout technique and suggest that device design plays a crucial role in creating a large orbital level spacing, which enables spin blockade in systems with strong electron correlation.
We have addressed the challenges associated with realizing few-electron states and rf reflectometry in ZnO quantum dots, which are essential for their application as qubits. 
Moreover, we have provided insights into the prospects for spin-state readout. 
These results pave the way for quantum information processing applications using ZnO quantum devices.

\section{Acknowledgements}
The authors thank K. Dezaki, M. Takeuchi, A. Kurita, and the RIEC Fundamental Technology Center and the Laboratory for Nanoelectronics and Spintronics for fruitful discussions and technical support.
Part of this work was supported by
MEXT Leading Initiative for Excellent Young Researchers, 
Grants-in-Aid for Scientific Research (21K18592, 22H04958, 23H01789, 23H04490), 
Tanigawa Foundation Research Grant, 
Maekawa Foundation Research Grant, 
The Foundation for Technology Promotion of Electronic Circuit Board, 
Iketani Science and Technology Foundation Research Grant, 
The Ebara Hatakeyama Memorial Foundation Research Grant, 
FRiD Tohoku University, 
and "Advanced Research Infrastructure for Materials and Nanotechnology in Japan (ARIM)" of the Ministry of Education, Culture, Sports, Science and Technology (MEXT) (Proposal Number JPMXP1224NM5072).
AIMR and MANA are supported by World Premier International Research Center Initiative (WPI), MEXT, Japan.

\section{Competing interests}
The authors declare no competing interests.

\bibliography{reference}

\end{document}